\documentclass[aps,prl,twocolumn,superscriptaddress]{revtex4-1}
\usepackage{graphicx}
\usepackage{dcolumn}
\usepackage{graphics}
\usepackage{hyperref}
\usepackage{float}
\usepackage{amsmath,amsfonts,amssymb}
\usepackage{physics}
\usepackage[normalem]{ulem}
\usepackage{xcolor}

\begin{document}
\title{Spin Qubits with Scalable milli-kelvin CMOS Control}

\author{Samuel K. Bartee}
\affiliation{ARC Centre of Excellence for Engineered Quantum Systems, School of Physics, The University of Sydney, Sydney, NSW 2006, Australia} 
\author{Will Gilbert}
\affiliation{School of Electrical Engineering and Telecommunications, University of New South Wales, Sydney, NSW 2052, Australia}
\affiliation{Diraq Pty. Ltd., Sydney, NSW, Australia}
\author{Kun Zuo}
\affiliation{ARC Centre of Excellence for Engineered Quantum Systems, School of Physics, The University of Sydney, Sydney, NSW 2006, Australia} 
\author{Kushal Das}
\affiliation{ARC Centre of Excellence for Engineered Quantum Systems, School of Physics, The University of Sydney, Sydney, NSW 2006, Australia}
\author{Tuomo Tanttu}
\affiliation{School of Electrical Engineering and Telecommunications, University of New South Wales, Sydney, NSW 2052, Australia}
\affiliation{Diraq Pty. Ltd., Sydney, NSW, Australia}
\author{Chih Hwan Yang}
\affiliation{School of Electrical Engineering and Telecommunications, University of New South Wales, Sydney, NSW 2052, Australia}
\affiliation{Diraq Pty. Ltd., Sydney, NSW, Australia}
\author{Nard Dumoulin Stuyck}
\affiliation{School of Electrical Engineering and Telecommunications, University of New South Wales, Sydney, NSW 2052, Australia}
\affiliation{Diraq Pty. Ltd., Sydney, NSW, Australia}
\author{Sebastian J. Pauka}
\affiliation{ARC Centre of Excellence for Engineered Quantum Systems, School of Physics, The University of Sydney, Sydney, NSW 2006, Australia}
\author{Rocky Y. Su}
\affiliation{School of Electrical Engineering and Telecommunications, University of New South Wales, Sydney, NSW 2052, Australia}
\author{Wee Han Lim}
\affiliation{School of Electrical Engineering and Telecommunications, University of New South Wales, Sydney, NSW 2052, Australia}
\affiliation{Diraq Pty. Ltd., Sydney, NSW, Australia}
\author{Santiago Serrano}
\affiliation{School of Electrical Engineering and Telecommunications, University of New South Wales, Sydney, NSW 2052, Australia}
\affiliation{Diraq Pty. Ltd., Sydney, NSW, Australia}
\author{Christopher C. Escott}
\affiliation{School of Electrical Engineering and Telecommunications, University of New South Wales, Sydney, NSW 2052, Australia}
\affiliation{Diraq Pty. Ltd., Sydney, NSW, Australia}
\author{Fay E. Hudson}
\affiliation{School of Electrical Engineering and Telecommunications, University of New South Wales, Sydney, NSW 2052, Australia}
\affiliation{Diraq Pty. Ltd., Sydney, NSW, Australia}
\author{Kohei M. Itoh}
\affiliation{School of Fundamental Science and Technology, Keio University, Kohoku-ku, Yokohama, Japan}
\author{Arne Laucht}
\affiliation{School of Electrical Engineering and Telecommunications, University of New South Wales, Sydney, NSW 2052, Australia}
\affiliation{Diraq Pty. Ltd., Sydney, NSW, Australia}
\author{Andrew S. Dzurak}
\affiliation{School of Electrical Engineering and Telecommunications, University of New South Wales, Sydney, NSW 2052, Australia}
\affiliation{Diraq Pty. Ltd., Sydney, NSW, Australia}
\author{David J. Reilly}
\email{David.Reilly@sydney.edu.au}
\affiliation{ARC Centre of Excellence for Engineered Quantum Systems, School of Physics, The University of Sydney, Sydney, NSW 2006, Australia}

\date{\today}

\begin{abstract}
A key virtue of spin qubits is their sub-micron footprint, enabling a single silicon chip to host the millions of qubits required to execute useful quantum algorithms with error correction \cite{Loss1998, maurand2016cmos, Veldhorst2017}. With each physical qubit needing multiple control lines however, a fundamental barrier to scale is the extreme density of connections that bridge quantum devices to their external control and readout hardware \cite{reilly2019, Vandersypen2017, degenhardt2017cmos}. A promising solution is to co-locate the control system proximal to the qubit platform at milli-kelvin temperatures, wired-up via miniaturized interconnects \cite{Pauka2021, reilly2015cryogenic, Charbon2018, geck2019control}. Even so, heat and crosstalk from closely integrated control have potential to degrade qubit performance, particularly for two-qubit entangling gates based on exchange coupling that are sensitive to electrical noise \cite{Dial2013, flyingqubits}. Here, we benchmark silicon MOS-style electron spin qubits controlled via heterogeneously-integrated cryo-CMOS circuits with a low enough power density to enable scale-up. Demonstrating that cryo-CMOS can efficiently enable universal logic operations for spin qubits, we go on to show that mill-kelvin control has little impact on the performance of single- and two-qubit gates. Given the complexity of our milli-kelvin CMOS platform, with some 100-thousand transistors, these results open the prospect of scalable control based on the tight packaging of spin qubits with a `chiplet style' control architecture.

\end{abstract}

\maketitle
\begin{figure*}[tp]
  \centering
  \includegraphics[scale = 0.8]{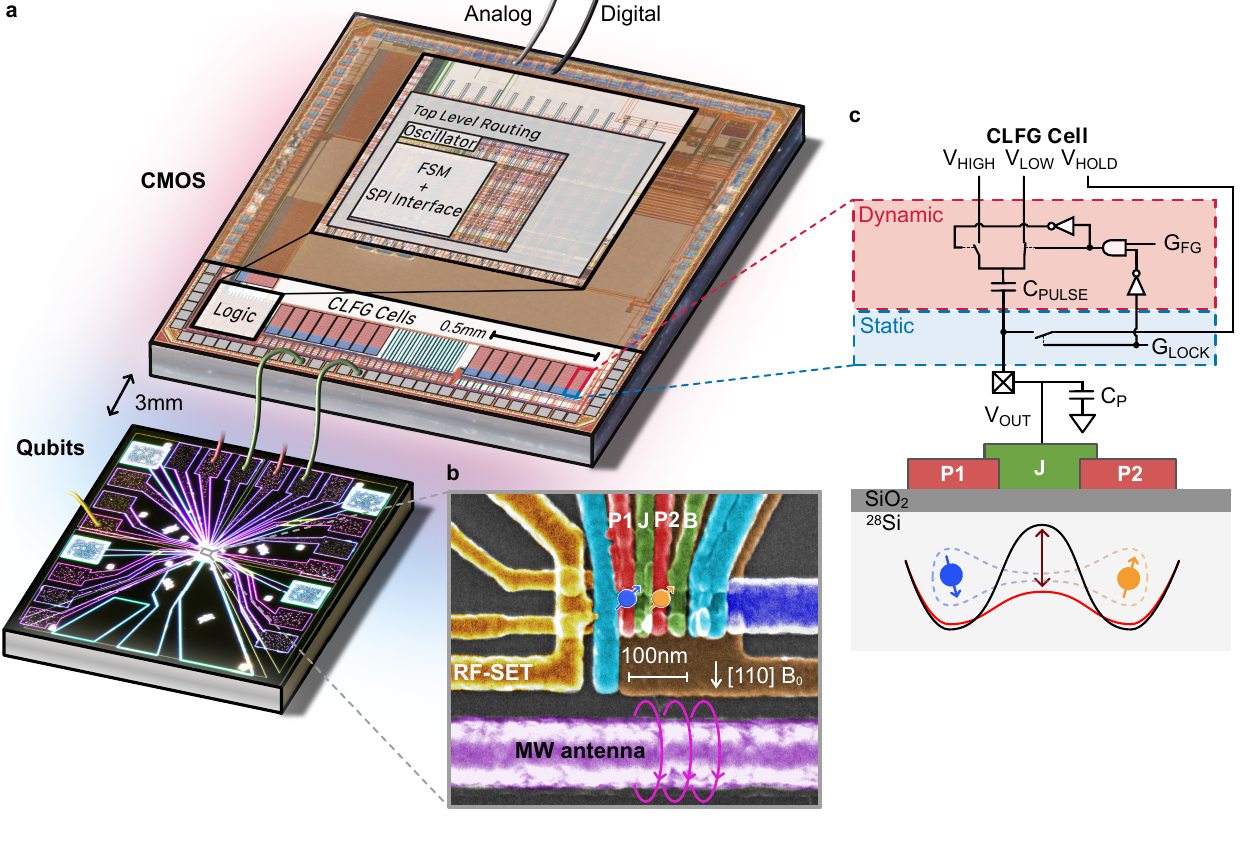}
  \caption{\textbf{Device and basic CMOS operation.} 
  \textbf{a}, The cryo-CMOS and the qubit chip are mounted on the same circuit board at mK temperatures and wire-bonded together. All other control systems are at room temperature.
  \textbf{b}, An electron micrograph  of a nominally identical silicon device to the one measured here. There are 32 CLFG cells on the chip, one of which is connected to gate $J$ that controls the coupling between the two quantum dots on the qubit chip. The other cell is connected to gate $B$, which acts as an additional barrier gate. All other fast-pulse gates ($P1$, $P2$, SET) and dc barrier gates are connected to room-temperature electronics. \textbf{c}, Schematic of a cell and its electrical connection to gate $J$ of the silicon device. Pulsing on this gate acts to modulate the tunnel coupling between the two quantum dots needed for single- and two-qubit control. 
  }
  \label{Fig1}
\end{figure*}

\noindent {\bf Introduction}\\
Utility scale quantum computing will likely require millions of physical qubits, operated via auxiliary classical systems that generate more than a trillion control signals per second \cite{beverland2022,reilly2015}. In realizing such a vast and complex platform, silicon qubits present advantages with their small footprint \cite{Kane1998}, long coherence times \cite{veldhorst2014addressable}, and inherent compatibility with very large scale integrated (VLSI) control circuits. While the potential for integrated control has been a key motivator for the progress in silicon-based qubits over the last two decades, to-date this aspect has remained largely undeveloped.

Despite the advantages of integrated control \cite{Vandersypen2017,reilly2019}, a serious concern arises from the heat and crosstalk generated by modern CMOS circuits. In relation to heat, this problem is eased by recent work showing that spin qubits continue to function at elevated temperatures \cite{Yang2020, Petit2020, huang2023}. Never-the-less, two-qubit entangling gates remain exquisitely sensitive to electrical noise \cite{Dial2013, flyingqubits}, arising for instance from volt-scale, sub-nanosecond switching of proximal CMOS transistors. One means of partially mitigating these adverse effects is to separate the control system to 4 kelvin, connecting to milli-kelvin qubits via long cables \cite{reilly2015cryogenic,vandersypen2021cmos}. Unfortunately, cable connectivity poses an additional barrier to scaling up the control interface \cite{reilly2019} given the extreme density of interconnects required to operate even modest numbers of qubits.

Here, we demonstrate the control of MOS-style silicon spin qubits using a heterogeneously-integrated cryo-CMOS chip operating at milli-kelvin temperatures, as shown in Fig.~\ref{Fig1}(a). Heterogeneous, `chiplet style' integration, as apposed to monolithic circuits, decouples the hot and noisy control system from sensitive qubits and retains the potential for dense, lithographically-defined chip-to-chip interconnects needed to manage the wiring challenge inherent to spin qubits. We demonstrate this chiplet architecture supports a control scheme that leverages a global resonance field to enable complete universal control of spin qubits using the baseband pulses that can be generated efficiently with proximal, low-power cryo-CMOS.

The details of the CMOS control chip have been reported previously with an early conceptual demonstration using GaAs quantum dot structures \cite{Pauka2021}. The effect of milli-kelvin CMOS on qubit performance however, has remained an open question until the present work. Naively, since the spin degree of freedom is somewhat decoupled from electrical noise, integrated CMOS is expected to have only a minor impact on single qubit operations. In contrast, coupling spins via Heisenberg exchange creates the most sensitive probe of voltage noise known \cite{Dial2013,flyingqubits} since the exchange energy can depend exponentially on gate voltage. Countering this intuition, we show that even for noise-sensitive two-qubit gates, our chiplet architecture, comprising some 100-thousand transistors, does not lead to a measurable reduction in coherence time. \\

\noindent {\bf Experimental Platform}\\
An electron micrograph of a silicon-metal-oxide-semiconductor (SiMOS) qubit device is shown in Fig.~\ref{Fig1}(b). The device is fabricated on an isotopically purified $^{28}$Si epilayer with residual $^{29}$Si concentration of 800~ppm \cite{Itoh2014}, and SiO$_2$ isolating layer with metal gates patterned in aluminium. Quantum dots hosting single spin qubits are formed under the plunger gates ($P1, P2$) at the Si/SiO$_2$ interface, and an exchange gate ($J$) modulates the tunnel coupling between the two dots, essential for two-qubit operations. A radio-frequency single-electron transistor (rf-SET) \cite{Angus2008} detects the charge state of the quantum dots on microsecond timescales by leveraging an off-chip $LC$ resonator operating near 400 MHz \cite{Reilly2007}, and a proximal microwave (MW) antenna generates an oscillating magnetic field for spin resonance control (see Fig.~\ref{Fig1}(b)).

The exchange gate $J$ and a barrier gate $B$ are both wire-bonded to the cryo-CMOS control chip \cite{Pauka2021} which is implemented in 28~nm fully depleted silicon-on-insulator technology (FDSOI). The chip contains a serial peripheral interface (SPI) to handle digital input instructions and a finite state machine (FSM) for on-chip digital logic. The FSM configures 32 analog `charge-lock fast-gate' (CLFG) circuit blocks, each of which can be used to control a gate electrode on the quantum device (see Fig.~\ref{Fig1}(c)). In this configuration, the charge is periodically stored and shuffled between small capacitors, leveraging the low leakage of transistors at cryogenic temperatures that maintain the potential during quantum operations. The cryo-CMOS chip also incorporates a ring oscillator and configurable register designed as a programmable internal trigger (see Extended Data Fig.~\ref{E3}(a) for oscillator schematics). Here, for convenience, we opt for external triggering.

The core functionality of a CLFG cell is to lock a static voltage bias and enable a fast pulse between two voltage levels, as outlined in Fig.~\ref{Fig1}(c). For instance, targeting gate $J$, the CLFG cell is programmed to first bring the gate to a potential $V_{\mathrm{OUT}}$, equal to the potential $V_{\mathrm{HOLD}}$ of an external source. Opening the switch $G_{\mathrm{LOCK}}$ under control of the FSM then `charge locks' this potential on the gate capacitor. Even though this floating capacitor is now galvanically disconnected from the source, a pulse can be induced by toggling the potential on the upper-plate of this capacitance between $V_{\mathrm{HIGH}}$ and $V_{\mathrm{LOW}}$, as shown in Fig.~\ref{Fig1}(c). This toggling is produced autonomously by the programmed on-chip FSM, leading to a modified output $V_{\mathrm{OUT}}$ by: $\Delta V_{\mathrm{PULSE}} = ({C_{\mathrm{PULSE}}}/{C_{\mathrm{P}} + C_{\mathrm{PULSE}}})\cdot(V_{\mathrm{HIGH}} - V_{\mathrm{LOW}})$, where $C_{\mathrm{P}}$ is the parasitic capacitance. This mechanism has previously been shown to produce pulse amplitudes of 100 mV at a power of $\sim$ 20 nW/MHz \cite{Pauka2021}. Below, we demonstrate how this architecture can be used to efficiently control spin qubits. \\

\begin{figure}
  \includegraphics[scale = 0.8]{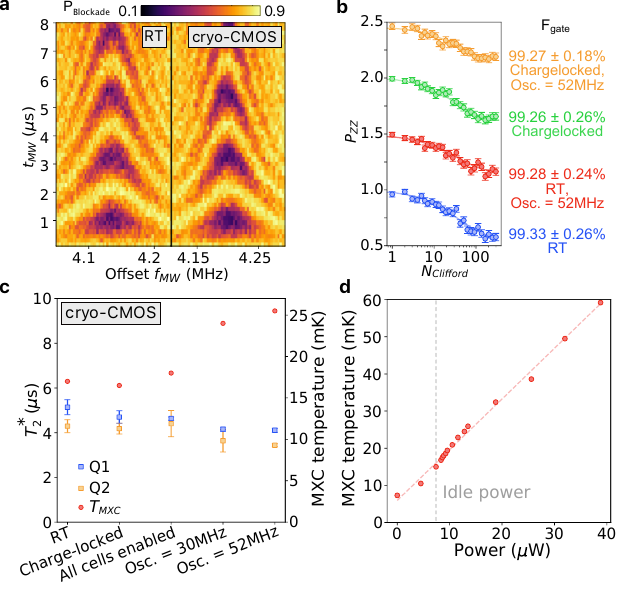}
  \caption{\textbf{Benchmarking single qubit cryo-CMOS performance}. 
  \textbf{a}, Single-qubit Rabi oscillations as a function of microwave frequency $f_{MW}$ and pulse time $t_{MW}$, performed with RT control. 
  \textbf{b}, Single qubit randomized benchmarking under various cryo-CMOS conditions. Traces are offset for clarity. 
  \textbf{c}, Single qubit $T_2^*$ coherence time as a function of select cryo-CMOS parameters. Unless otherwise indicated, all data taken here uses cryo-CMOS control.  \textbf{d}, Mixing chamber temperature with cryo-CMOS power.}
  \label{Fig2}
\end{figure}

\noindent {\bf Experimental Results and Demonstrations}\\
Turning to evaluate single qubit gates, we first establish a baseline using all room-temperature (RT) electronics for control. Following the usual protocol for two-spin manipulation \cite{Loss1998,petta2005,martins2016}, the singlet state is first prepared in the (1,3) charge configuration using pulses applied to detuning gates $P1$ and $P2$, [with (n,m) labeling the number of electrons in each dot under $P1$ and $P2$ respectively]. A pulse applied to the $J$-gate, connected to $V_{\mathrm{HOLD}}$, then increases the barrier, separating the two electrons into each dot where they are independently addressed using the MW antenna via their unique resonance frequency ($f_{ESR}$ = 13.9~GHz for a field $B_0$ = 0.5~T). Free-induction decay (FID) of the target spin is produced by applying MW power to the on-chip ESR line. Finally, a second pulse of the $J$-gate returns the spins to the readout configuration where Pauli spin blockade (PSB) enables spin-to-charge conversion \cite{psbono, psbpetta} and measurement via the rf-SET. The shot-averaged readout signal as a function of MW power and frequency is shown in Fig.~\ref{Fig2}(a). Beyond FID, we further establish our RT baseline by performing pulse sequences implementing Hahn echo (to measure coherence time $T_2$, see Extended Data Fig.~\ref{E4}) and randomized benchmarking (to measure qubit control fidelity) \cite{RBM1,RBM2}. 

In this single-qubit measurement, the function of the $J$-gate pulse is to separate the two-spin system for controlled rotation via spin resonance. As such, electrical noise, coupled via the $J$-gate or other means is unlikely to impact qubit fidelity in the limit that the pulse amplitude and duration are sufficiently large to fully separate the spins. Even so, we now evaluate the impact of cryo-CMOS control on single qubit performance by performing the same protocol outlined above, but now with charge-locking applied to the $J$-gate and the pulse produced using a CLFG cell under control of the FSM. Again we generate FID data and quantitatively compare between CMOS and RT control using randomized benchmarking protocols, as shown in Fig.~\ref{Fig2}(b). A slight degradation in qubit fidelity is observed (0.07\%), likely due to unmitigated heat from the CMOS. We discuss heating in detail below.

Although electrical noise on the $J$-gate does not directly couple to single spins, heat and drift in gate potential over longer timescales can impact qubit performance. Gate noise can also dc-Stark shift the qubit frequency in certain regimes (discussed further below). To investigate these mechanisms we extract the time-ensemble average coherence time $T_2^*$ for each qubit, repeatedly measured as each circuit block of the CMOS chip is powered-up. Comparing this data in Fig.~\ref{Fig2}(c) again shows a small impact with respect to our RT baseline, correlating with a slight rise in the base temperature of the refrigerator (see figure caption for details). 

A drawback of the control scheme outlined above is its reliance on qubit-specific microwave pulses, which, despite cryo-CMOS gate control, still requires additional RT pulse generators (and cables) for each qubit. We next demonstrate an alternate control approach that leverages a continuous wave (CW) global microwave field, sourced from RT but common to all qubits \cite{Kane1998,Vahapoglu_2022}. Key to this scheme is the ability to tune the spin resonance frequency of a qubit using a gate voltage \cite{laucht2017dressed}. This dc Stark shift enables a gate pulse to bring a qubit into resonance with the global field for a controlled amount of time to produce a rotation in the qubit state vector. With a single microwave tone from RT, cryo-CMOS produces the `baseband' gate pulses that independently bring each qubit into and out of resonance with the global field. 

\begin{figure}
  \includegraphics[scale=0.8]{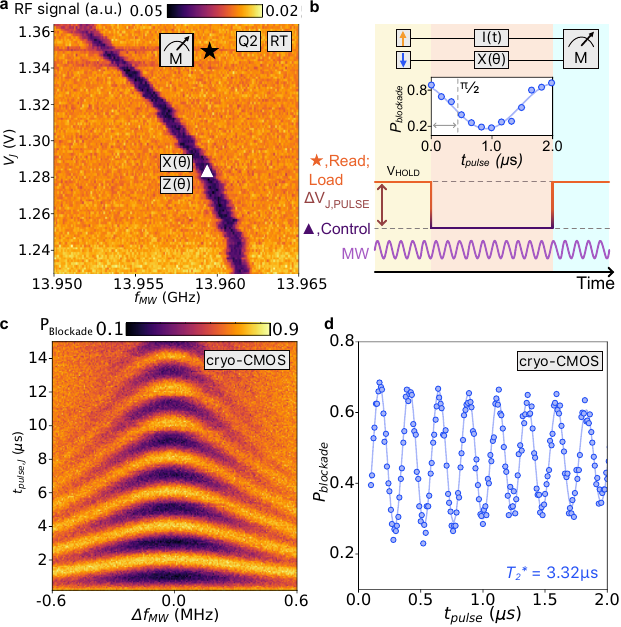}
  \caption{\textbf{Single qubit gates with cryo-CMOS.}
    \textbf{a}, ESR spectra as a function of microwave frequency and $J$ gate voltage. $Q_2$ exhibits a significant Stark shift, which allows for on-resonance single qubit operations (X, Z; $\blacktriangle$), and off-resonance loading and measurement (M; $\star$).
    \textbf{b}, Schematic of the pulsing sequence when using effective global control. Qubit rotations are determined by $J$ pulse time, with the MW tone fixed to the maximum $t_{pulse}$ time, extending into load and read stages. 
    \textbf{c}, Rabi chevron of $Q_2$ and
    \textbf{d}, Ramsey coherence time, using global control and cryo-CMOS pulsing at $B_0 = 0.5$~T.}
  \label{Fig3}
\end{figure}

The global-MW scheme requires a calibration of the unique dc-Stark shift produced by the $J$-gate on say, qubit $Q_2$, as shown in Fig.~\ref{Fig3}(a). The sizable shift ($\sim$ 1~MHz/10~mV) is well-matched to the voltage pulses that can be efficiently generated with proximal low-power CMOS. Here, the spins are initialized in the ($T^-$) ground state via off-resonance relaxation, with the $J$-gate pulse producing the time-controlled Stark shift. Repeating this sequence as a function of pulse length yields the coherent oscillations shown in Fig.~\ref{Fig3}(c). For this measurement, the width of the pulse is set by the timing of the trigger fed to the CMOS from RT. Conceptually, this reliance on RT triggering may appear to be a limitation of our CMOS circuits, however, we note that fine time resolution is only needed to map out coherent oscillations. In contrast, once calibrated, logic gates require a fixed time pulse, for instance, 1.034~$\mu$s to produce a $\pi/2$ rotation. As such, these fixed time pulses are straightforward to implement with our CMOS architecture.

\begin{figure}
  \includegraphics[scale=0.8]{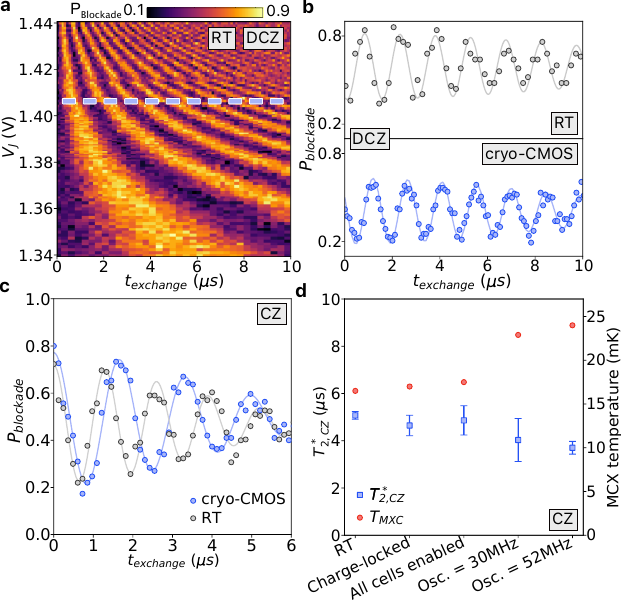}
  \caption{\textbf{Two-qubit operations using cryo-CMOS.} 
  \textbf{a}, Decoupled controlled phase (DCZ) oscillations as a function of exchange time $t_{exchange}$ and $V_J$, performed using RT control.
  \textbf{b}, RT and Cryo-CMOS-enabled DCZ oscillations at a set $V_J$, indicated in a. Set level is determined by $V_{HOLD}$, which can be tuned for stronger qubit interaction. 
  \textbf{c}, Comparison of visibility between RT and cryo-CMOS control using identical pulsing methods on all gates. Increased state preparation and measurement error is present due to two-level-only approach to pulsing J.
  \textbf{d}, CZ coherence times of our two-qubit control conditions under various cryo-CMOS parameters. $J$ pulses are generated by cryo-CMOS unless otherwise indicated.}
  \label{Fig4}
  
\end{figure}

Lastly we turn to evaluate two-qubit logic gates, which provide the most stringent test for crosstalk or electrical noise stemming from proximal milli-kelvin CMOS control. Here, the target qubit is rotated about the $Z$-axis depending on the state of the control qubit, with the gate-tunable exchange interaction modulating the coupling between the two electrons \cite{Loss1998, petta2005}. As a baseline we first perform a decoupled controlled phase gate (DCZ) using all RT instrumentation. The DCZ gate 
incorporates a spin-echo sequence with coherent rotation about the $Z$-axis of angle $\phi = J(\epsilon)t_{ex}\hbar$, enabled by turning-on exchange for a controlled time with $J$-gate pulse of duration $t_{ex}$ (see Extended Data Fig.~\ref{E2}(d)). The resulting readout probability with $J$-gate pulse width is shown in Fig.~\ref{Fig4}(a).  The DCZ gate is sensitive to high-frequency electrical noise arising either directly from exchange-gate voltage fluctuations or, indirectly, from noise in the (gradient) magnetic field or variation in g-factors from gate-induced movement in position of the electron wavefunction. 

Comparing now cryo-CMOS control to our RT baseline, we observe that the coherent exchange oscillations shows similar behaviour, as shown in Fig.~\ref{Fig4}(b). These results immediately confirm the utility of proximal milli-kelvin CMOS for controlling two-qubit logic gates. Close inspection perhaps suggests a suppression in visibility for the CMOS data, however this can be explained by the available resolution during readout and preparation phase, as opposed to during control.

As a noise diagnostic tool, it is also worth noting that the DCZ gate is somewhat limited since the spin-echo sequence decouples the spin dynamics from low frequency noise. Removing the echo pulses then opens the bandwidth to now include all of the low frequency components down to quasi-dc, offering a better measure of the total aggregate noise inherent in the system. A comparison between RT control and cryo-CMOS is made in the FID data shown in Fig.~\ref{Fig4}(c), now without spin-echo. These datasets constitute a measure of the ensemble average coherence time associated with the exchange gate, $T^*_{2, CZ}$. Lastly, using this parameter as a wideband measure of noise, Fig.~\ref{Fig4}(d) compares $T^*_{2, CZ}$ for RT control, cryo-CMOS with a single charge-locked cell, all cells locked (mirroring $J$-gate pulses), and as a function of CMOS oscillator frequency. A slight reduction in $T^*_{2, CZ}$ ($\sim$ 20\%) is observed at the highest clock frequencies, which, given the slight corresponding increase in refrigerator temperature, likely stems from parasitic heating. \\

\noindent {\bf Discussion}\\
Our cryo-CMOS control chip comprises complex mixed-signal circuits realized using more than 100-thousand transistors. The vast majority of these are used in the digital sub-systems and related circuit blocks, accounting for a fixed overhead power of 10s of microWatts. On top of this constant offset power from the digital blocks, the CLFG analog cells each contribute $\sim$20 nW/MHz when generating 100 mV amplitude pulses, enabling many thousands of cells (and thus gate pulses) to fit within the cooling budget of a commercial dilution refrigerator ($\sim$ 1 mW at 100 mK) \cite{Pauka2021}. Aside from the cooling limits of the refrigerator however, a challenge arises in the thermal management of hot control systems to ensure the routing of heat bypasses proximal, cold quantum devices. Here, we have made little attempt to mitigate this parasitic heating, simply wire-bonding the chips together in a standard package. This arrangement can lead to elevated electron temperatures in the quantum device (see Extended Data Fig.~\ref{E3}) even when the refrigerator remains cold and is the likely explanation for the small impact we observe in qubit fidelity when the largest CMOS circuits are powered up at the highest clock rates. As such, we emphasize that there is significant opportunity to suppress parasitic heating by employing separate parallel cooling pathways for the CMOS chip and quantum plane \cite{Pauka2021}. The use of heterogeneous, rather than monolithic integration opens new thermal configuration options in this regard.

Beyond direct heating, the close presence of 100-thousand transistors, with volt-scale biasing and sub-nanosecond rise and fall times, can create an exceedingly noisy environment in which to operate electrically sensitive qubits. It is perhaps surprising then that we observe the CMOS chip to have only a small impact on qubit performance. Even further, the small degradation in fidelity is likely explained entirely from parasitic heating, rather than from electrical noise. Certainly, our use of CMOS design-rules that minimize external crosstalk are important, however beyond these, we suggest three additional aspects that likely reduce electrical noise. Firstly, since the physical temperature of the CMOS die is a few-hundred milli-kelvin, thermal noise contributions are significantly suppressed. Also, the chip-to-chip interconnect likely has a relatively low bandwidth, filtering noise above a few giga-Hertz. Lastly, we note that the action of the CLFG circuits effectively decouple the CMOS from the quantum device when in charge-lock mode, except for a very small coupling capacitor. Taken collectively these aspects further underscore the utility of heterogeneous over monolithic integration for mitigating crosstalk and heating. In addition to addressing the challenges posed by scaling-up qubits, cryo-CMOS using a chiplet architecture may also prove useful in generating ultra-fast, low thermal noise control pulses that probe fundamental physics in mesoscale quantum devices \cite{Glattli}. 

In conclusion, the results presented here demonstrate the viability of heterogeneous, milli-kelvin CMOS for scalable control of spin qubits. Beyond addressing the interconnect bottleneck posed by cryogenic qubit platforms, these results show that degradation in qubit performance from  milli-kelvin CMOS is very limited. Although our focus here has been controlling spin qubits based on single electrons, we draw attention to the inherent compatibility of our control architecture with other flavours of spin qubits, for instance, exchange only qubits that leverage square voltage pulses exclusively \cite{Weinstein2023}.

\section*{Methods}
\setcounter{subsection}{0}
\subsection{Measurement Setup}
Measurements are performed in a Bluefors LD400 dilution refrigerator with a base temperature of 7 mK. The qubit chip and cryo-CMOS chip are packaged on the same FR4 printed circuit board (PCB) \cite{collessRSI}, separated by 3~mm and wirebonded together. The daughterboard PCB is placed on a custom motherboard and electrically connected via an interposer. The motherboard manages signals from room temperature and the ESR line connects directly to the daughterboard via a miniSMP. The PCB setup is mounted in a magnetic field, on a cold finger. The external magnetic field is supplied by an Oxford MercuryiPS-M magnet power supply. DC voltages are generated by an in-house custom digitial to analog converter (DAC). A Quantum Machines OPX+ generates RT dynamic pulses, the 400~MHz signal for rf-readout, I/Q and pulse modulation for the MW source as well as trigger lines to the cryo-CMOS chip and MW source. Dynamic and dc voltage sources are combined at room temperature using custom voltage combiners. The OPX has a sampling rate of 1~GS/s and a clock rate of 250~MHz. The MW tone is generated by a Keysight PSG E8267D Vector Signal Generator with a signal spanning 250~kHz to 31.8~GHz.
We use an rf-SET and rf reflectometry readout, comprising a Low Noise Factory LNF-LNC0.3-14A amplifier at the 4~K stage of the refrigerator and a Minicircuits ZX60-P103LN+ at RT for signal amplification. The directional coupler in Fig.~\ref{Fig1}(a) is a Minicircuits ZEDC-10-182-S+ 10-1800~MHz.\\

\subsection{Cryo-CMOS Programming}
Our cryo-CMOS receives programming instructions, power, dc bias, and dynamic voltage levels from an in-house RT DAC \cite{waddy2023} and is configured to receive an external trigger from the OPX+, as well as RT dynamic pulses for passthrough RT control. Some of the programming instructions as well as the external trigger are received by the same input line. Appropriate input is handled by a Minicircuits RC-4SPDT-A18 DC-18~GHz rf switch, which takes inputs from both the DAC and OPX+. The rf switch is programmed to work in unison with charge-locking commands sent to the cryo-CMOS, switching inputs from the DAC to the OPX+ once programming is complete.

Due to the low leakage rate of our cryo-CMOS transistors, there are no charge-lock refresh cycles of our charge-locked gate when performing experiments. Maximum experiment times are approximately one hour, usually single-qubit RBM or PSD measurements.

\section*{Acknowledgements}
We thank Y. Yang and R. Kalra for technical contributions and  discussions. This research was supported by Microsoft Corporation (CPD 1-4) and by the Australian Research Council Centre of Excellence for Engineered Quantum Systems (EQUS, CE170100009). We acknowledge support from the Australian Research Council (FL190100167 and CE170100012), the US Army Research Office (W911NF-23-10092), the US Air Force Office of Scientific Research (FA2386-22-1-4070) and the NSW Node of the Australian National Fabrication Facility. The views and conclusions contained in this document are those of the authors and should not be interpreted as representing the official policies, either expressed or implied, of Microsoft Corporation, the Army Research Office, the US Air Force or the US government. The US government is authorized to reproduce and distribute reprints for government purposes notwithstanding any copyright notation herein. R.Y.S. and S.S. acknowledge support from the Sydney Quantum Academy.






\bibliographystyle{naturemag}
\bibliography{references}

\begin{thebibliography}{10}
\expandafter\ifx\csname url\endcsname\relax
  \def\url#1{\texttt{#1}}\fi
\expandafter\ifx\csname urlprefix\endcsname\relax\def\urlprefix{URL }\fi
\providecommand{\bibinfo}[2]{#2}
\providecommand{\eprint}[2][]{\url{#2}}

\bibitem{Loss1998}
\bibinfo{author}{Loss, D.} \& \bibinfo{author}{DiVincenzo, D.~P.}
\newblock \bibinfo{title}{Quantum computation with quantum dots}.
\newblock \emph{\bibinfo{journal}{Phys. Rev. A}} \textbf{\bibinfo{volume}{57}},
  \bibinfo{pages}{120--126} (\bibinfo{year}{1998}).
\newblock \urlprefix\url{https://link.aps.org/doi/10.1103/PhysRevA.57.120}.

\bibitem{maurand2016cmos}
\bibinfo{author}{Maurand, R.} \emph{et~al.}
\newblock \bibinfo{title}{A cmos silicon spin qubit}.
\newblock \emph{\bibinfo{journal}{Nature communications}}
  \textbf{\bibinfo{volume}{7}}, \bibinfo{pages}{13575} (\bibinfo{year}{2016}).

\bibitem{Veldhorst2017}
\bibinfo{author}{Veldhorst, M.}, \bibinfo{author}{Eenink, H. G.~J.},
  \bibinfo{author}{Yang, C.~H.} \& \bibinfo{author}{Dzurak, A.~S.}
\newblock \bibinfo{title}{{Silicon CMOS architecture for a spin-based quantum
  computer}}.
\newblock \emph{\bibinfo{journal}{Nature Communications}}
  \textbf{\bibinfo{volume}{8}}, \bibinfo{pages}{1766} (\bibinfo{year}{2017}).
\newblock \urlprefix\url{https://doi.org/10.1038/s41467-017-01905-6}.

\bibitem{reilly2019}
\bibinfo{author}{Reilly, D.}
\newblock \bibinfo{title}{Challenges in scaling-up the control interface of a
  quantum computer}.
\newblock In \emph{\bibinfo{booktitle}{2019 IEEE International Electron Devices
  Meeting (IEDM)}}, \bibinfo{pages}{31--7} (\bibinfo{organization}{IEEE},
  \bibinfo{year}{2019}).

\bibitem{Vandersypen2017}
\bibinfo{author}{Vandersypen, L. M.~K.} \emph{et~al.}
\newblock \bibinfo{title}{{Interfacing spin qubits in quantum dots and
  donors—hot, dense, and coherent}}.
\newblock \emph{\bibinfo{journal}{npj Quantum Information}}
  \textbf{\bibinfo{volume}{3}}, \bibinfo{pages}{34} (\bibinfo{year}{2017}).
\newblock \urlprefix\url{https://doi.org/10.1038/s41534-017-0038-y}.

\bibitem{degenhardt2017cmos}
\bibinfo{author}{Degenhardt, C.}, \bibinfo{author}{Geck, L.},
  \bibinfo{author}{Kruth, A.}, \bibinfo{author}{Vliex, P.} \&
  \bibinfo{author}{van Waasen, S.}
\newblock \bibinfo{title}{Cmos based scalable cryogenic control electronics for
  qubits}.
\newblock In \emph{\bibinfo{booktitle}{2017 IEEE International Conference on
  Rebooting Computing (ICRC)}}, \bibinfo{pages}{1--4}
  (\bibinfo{organization}{IEEE}, \bibinfo{year}{2017}).

\bibitem{Pauka2021}
\bibinfo{author}{Pauka, S.~J.} \emph{et~al.}
\newblock \bibinfo{title}{{A cryogenic CMOS chip for generating control signals
  for multiple qubits}}.
\newblock \emph{\bibinfo{journal}{Nature Electronics}}
  \textbf{\bibinfo{volume}{4}}, \bibinfo{pages}{64--70} (\bibinfo{year}{2021}).
\newblock \urlprefix\url{https://doi.org/10.1038/s41928-020-00528-y}.

\bibitem{reilly2015cryogenic}
\bibinfo{author}{Hornibrook, J.} \emph{et~al.}
\newblock \bibinfo{title}{Cryogenic control architecture for large-scale
  quantum computing}.
\newblock \emph{\bibinfo{journal}{Physical Review Applied}}
  \textbf{\bibinfo{volume}{3}}, \bibinfo{pages}{024010} (\bibinfo{year}{2015}).

\bibitem{Charbon2018}
\bibinfo{author}{Patra, B.} \emph{et~al.}
\newblock \bibinfo{title}{Cryo-cmos circuits and systems for quantum computing
  applications}.
\newblock \emph{\bibinfo{journal}{IEEE Journal of Solid-State Circuits}}
  \textbf{\bibinfo{volume}{53}}, \bibinfo{pages}{309--321}
  (\bibinfo{year}{2018}).

\bibitem{geck2019control}
\bibinfo{author}{Geck, L.}, \bibinfo{author}{Kruth, A.},
  \bibinfo{author}{Bluhm, H.}, \bibinfo{author}{van Waasen, S.} \&
  \bibinfo{author}{Heinen, S.}
\newblock \bibinfo{title}{Control electronics for semiconductor spin qubits}.
\newblock \emph{\bibinfo{journal}{Quantum science and technology}}
  \textbf{\bibinfo{volume}{5}}, \bibinfo{pages}{015004} (\bibinfo{year}{2019}).

\bibitem{Dial2013}
\bibinfo{author}{Dial, O.~E.} \emph{et~al.}
\newblock \bibinfo{title}{Charge noise spectroscopy using coherent exchange
  oscillations in a singlet-triplet qubit}.
\newblock \emph{\bibinfo{journal}{Phys. Rev. Lett.}}
  \textbf{\bibinfo{volume}{110}}, \bibinfo{pages}{146804}
  (\bibinfo{year}{2013}).
\newblock
  \urlprefix\url{https://link.aps.org/doi/10.1103/PhysRevLett.110.146804}.

\bibitem{flyingqubits}
\bibinfo{author}{Thiney, V.} \emph{et~al.}
\newblock \bibinfo{title}{In-flight detection of few electrons using a
  singlet-triplet spin qubit}.
\newblock \emph{\bibinfo{journal}{Physical Review Research}}
  \textbf{\bibinfo{volume}{4}}, \bibinfo{pages}{043116} (\bibinfo{year}{2022}).

\bibitem{beverland2022}
\bibinfo{author}{Beverland, M.~E.} \emph{et~al.}
\newblock \bibinfo{title}{Assessing requirements to scale to practical quantum
  advantage (2022)}.
\newblock \emph{\bibinfo{journal}{arXiv preprint arXiv:2211.07629}}
  (\bibinfo{year}{2022}).

\bibitem{reilly2015}
\bibinfo{author}{Reilly, D.~J.}
\newblock \bibinfo{title}{Engineering the quantum-classical interface of
  solid-state qubits}.
\newblock \emph{\bibinfo{journal}{npj Quantum Information}}
  \textbf{\bibinfo{volume}{1}}, \bibinfo{pages}{1--10} (\bibinfo{year}{2015}).

\bibitem{Kane1998}
\bibinfo{author}{Kane, B.~E.}
\newblock \bibinfo{title}{{A silicon-based nuclear spin quantum computer}}.
\newblock \emph{\bibinfo{journal}{Nature}} \textbf{\bibinfo{volume}{393}},
  \bibinfo{pages}{133--137} (\bibinfo{year}{1998}).
\newblock \urlprefix\url{https://doi.org/10.1038/30156}.

\bibitem{veldhorst2014addressable}
\bibinfo{author}{Veldhorst, M.} \emph{et~al.}
\newblock \bibinfo{title}{An addressable quantum dot qubit with fault-tolerant
  control-fidelity}.
\newblock \emph{\bibinfo{journal}{Nature nanotechnology}}
  \textbf{\bibinfo{volume}{9}}, \bibinfo{pages}{981--985}
  (\bibinfo{year}{2014}).

\bibitem{Yang2020}
\bibinfo{author}{Yang, C.~H.} \emph{et~al.}
\newblock \bibinfo{title}{{Operation of a silicon quantum processor unit cell
  above one kelvin}}.
\newblock \emph{\bibinfo{journal}{Nature}} \textbf{\bibinfo{volume}{580}},
  \bibinfo{pages}{350--354} (\bibinfo{year}{2020}).
\newblock \urlprefix\url{https://doi.org/10.1038/s41586-020-2171-6}.

\bibitem{Petit2020}
\bibinfo{author}{Petit, L.} \emph{et~al.}
\newblock \bibinfo{title}{{Universal quantum logic in hot silicon qubits}}.
\newblock \emph{\bibinfo{journal}{Nature}} \textbf{\bibinfo{volume}{580}},
  \bibinfo{pages}{355--359} (\bibinfo{year}{2020}).
\newblock \urlprefix\url{https://doi.org/10.1038/s41586-020-2170-7}.

\bibitem{huang2023}
\bibinfo{author}{Huang, J.~Y.} \emph{et~al.}
\newblock \bibinfo{title}{High-fidelity spin qubit operation and algorithmic
  initialization above 1 k}.
\newblock \emph{\bibinfo{journal}{Nature}} \textbf{\bibinfo{volume}{627}},
  \bibinfo{pages}{772--777} (\bibinfo{year}{2024}).

\bibitem{vandersypen2021cmos}
\bibinfo{author}{Xue, X.} \emph{et~al.}
\newblock \bibinfo{title}{Cmos-based cryogenic control of silicon quantum
  circuits}.
\newblock \emph{\bibinfo{journal}{Nature}} \textbf{\bibinfo{volume}{593}},
  \bibinfo{pages}{205--210} (\bibinfo{year}{2021}).

\bibitem{Itoh2014}
\bibinfo{author}{Itoh, K.~M.} \& \bibinfo{author}{Watanabe, H.}
\newblock \bibinfo{title}{{Isotope engineering of silicon and diamond for
  quantum computing and sensing applications}}.
\newblock \emph{\bibinfo{journal}{MRS Communications}}
  \textbf{\bibinfo{volume}{4}}, \bibinfo{pages}{143--157}
  (\bibinfo{year}{2014}).
\newblock \urlprefix\url{https://doi.org/10.1557/mrc.2014.32}.

\bibitem{Angus2008}
\bibinfo{author}{Angus, S.~J.}, \bibinfo{author}{Ferguson, A.~J.},
  \bibinfo{author}{Dzurak, A.~S.} \& \bibinfo{author}{Clark, R.~G.}
\newblock \bibinfo{title}{{A silicon radio-frequency single electron
  transistor}}.
\newblock \emph{\bibinfo{journal}{Applied Physics Letters}}
  \textbf{\bibinfo{volume}{92}}, \bibinfo{pages}{112103}
  (\bibinfo{year}{2008}).
\newblock \urlprefix\url{https://doi.org/10.1063/1.2831664}.

\bibitem{Reilly2007}
\bibinfo{author}{Reilly, D.~J.}, \bibinfo{author}{Marcus, C.~M.},
  \bibinfo{author}{Hanson, M.~P.} \& \bibinfo{author}{Gossard, A.~C.}
\newblock \bibinfo{title}{{Fast single-charge sensing with a rf quantum point
  contact}}.
\newblock \emph{\bibinfo{journal}{Applied Physics Letters}}
  \textbf{\bibinfo{volume}{91}}, \bibinfo{pages}{162101}
  (\bibinfo{year}{2007}).
\newblock \urlprefix\url{https://doi.org/10.1063/1.2794995}.

\bibitem{petta2005}
\bibinfo{author}{Petta, J.~R.} \emph{et~al.}
\newblock \bibinfo{title}{Coherent manipulation of coupled electron spins in
  semiconductor quantum dots}.
\newblock \emph{\bibinfo{journal}{Science}} \textbf{\bibinfo{volume}{309}},
  \bibinfo{pages}{2180--2184} (\bibinfo{year}{2005}).

\bibitem{martins2016}
\bibinfo{author}{Martins, F.} \emph{et~al.}
\newblock \bibinfo{title}{Noise suppression using symmetric exchange gates in
  spin qubits}.
\newblock \emph{\bibinfo{journal}{Physical review letters}}
  \textbf{\bibinfo{volume}{116}}, \bibinfo{pages}{116801}
  (\bibinfo{year}{2016}).

\bibitem{psbono}
\bibinfo{author}{Ono, K.}, \bibinfo{author}{Austing, D.},
  \bibinfo{author}{Tokura, Y.} \& \bibinfo{author}{Tarucha, S.}
\newblock \bibinfo{title}{Current rectification by pauli exclusion in a weakly
  coupled double quantum dot system}.
\newblock \emph{\bibinfo{journal}{Science}} \textbf{\bibinfo{volume}{297}},
  \bibinfo{pages}{1313--1317} (\bibinfo{year}{2002}).

\bibitem{psbpetta}
\bibinfo{author}{Johnson, A.~C.}, \bibinfo{author}{Petta, J.~R.},
  \bibinfo{author}{Marcus, C.~M.}, \bibinfo{author}{Hanson, M.~P.} \&
  \bibinfo{author}{Gossard, A.~C.}
\newblock \bibinfo{title}{Singlet-triplet spin blockade and charge sensing in a
  few-electron double quantum dot}.
\newblock \emph{\bibinfo{journal}{Phys. Rev. B}} \textbf{\bibinfo{volume}{72}},
  \bibinfo{pages}{165308} (\bibinfo{year}{2005}).
\newblock \urlprefix\url{https://link.aps.org/doi/10.1103/PhysRevB.72.165308}.

\bibitem{RBM1}
\bibinfo{author}{Knill, E.} \emph{et~al.}
\newblock \bibinfo{title}{Randomized benchmarking of quantum gates}.
\newblock \emph{\bibinfo{journal}{Phys. Rev. A}} \textbf{\bibinfo{volume}{77}},
  \bibinfo{pages}{012307} (\bibinfo{year}{2008}).
\newblock \urlprefix\url{https://link.aps.org/doi/10.1103/PhysRevA.77.012307}.

\bibitem{RBM2}
\bibinfo{author}{Magesan, E.}, \bibinfo{author}{Gambetta, J.~M.} \&
  \bibinfo{author}{Emerson, J.}
\newblock \bibinfo{title}{Characterizing quantum gates via randomized
  benchmarking}.
\newblock \emph{\bibinfo{journal}{Phys. Rev. A}} \textbf{\bibinfo{volume}{85}},
  \bibinfo{pages}{042311} (\bibinfo{year}{2012}).
\newblock \urlprefix\url{https://link.aps.org/doi/10.1103/PhysRevA.85.042311}.

\bibitem{Vahapoglu_2022}
\bibinfo{author}{Vahapoglu, E.} \emph{et~al.}
\newblock \bibinfo{title}{Coherent control of electron spin qubits in silicon
  using a global field}.
\newblock \emph{\bibinfo{journal}{npj Quantum Information}}
  \textbf{\bibinfo{volume}{8}} (\bibinfo{year}{2022}).
\newblock \urlprefix\url{https://doi.org/10.1038%2Fs41534-022-00645-w}.

\bibitem{laucht2017dressed}
\bibinfo{author}{Laucht, A.} \emph{et~al.}
\newblock \bibinfo{title}{A dressed spin qubit in silicon}.
\newblock \emph{\bibinfo{journal}{Nature nanotechnology}}
  \textbf{\bibinfo{volume}{12}}, \bibinfo{pages}{61--66}
  (\bibinfo{year}{2017}).

\bibitem{Glattli}
\bibinfo{author}{Dubois, J.} \emph{et~al.}
\newblock \bibinfo{title}{Minimal-excitation states for electron quantum optics
  using levitons}.
\newblock \emph{\bibinfo{journal}{Nature}} \textbf{\bibinfo{volume}{502}},
  \bibinfo{pages}{659--663} (\bibinfo{year}{2013}).

\bibitem{Weinstein2023}
\bibinfo{author}{Weinstein, A.~J.} \emph{et~al.}
\newblock \bibinfo{title}{{Universal logic with encoded spin qubits in
  silicon}}.
\newblock \emph{\bibinfo{journal}{Nature}} \textbf{\bibinfo{volume}{615}},
  \bibinfo{pages}{817--822} (\bibinfo{year}{2023}).
\newblock \urlprefix\url{https://doi.org/10.1038/s41586-023-05777-3}.

\bibitem{collessRSI}
\bibinfo{author}{Colless, J.} \& \bibinfo{author}{Reilly, D.}
\newblock \bibinfo{title}{Modular cryogenic interconnects for multi-qubit
  devices}.
\newblock \emph{\bibinfo{journal}{Review of Scientific Instruments}}
  \textbf{\bibinfo{volume}{85}} (\bibinfo{year}{2014}).

\bibitem{waddy2023}
\bibinfo{author}{Waddy, S.}
\newblock \emph{\bibinfo{title}{Quantum Readout and Control: Scalable Hardware
  and Techniques for Solid State Quantum Systems}}.
\newblock Ph.D. thesis, \bibinfo{address}{Sydney, NSW} (\bibinfo{year}{2023}).

\bibitem{Sarma2007}
\bibinfo{author}{Cywi\ifmmode~\acute{n}\else \'{n}\fi{}ski, L.},
  \bibinfo{author}{Lutchyn, R.~M.}, \bibinfo{author}{Nave, C.~P.} \&
  \bibinfo{author}{Das~Sarma, S.}
\newblock \bibinfo{title}{How to enhance dephasing time in superconducting
  qubits}.
\newblock \emph{\bibinfo{journal}{Phys. Rev. B}} \textbf{\bibinfo{volume}{77}},
  \bibinfo{pages}{174509} (\bibinfo{year}{2008}).
\newblock \urlprefix\url{https://link.aps.org/doi/10.1103/PhysRevB.77.174509}.

\bibitem{Alvarez2011}
\bibinfo{author}{\'Alvarez, G.~A.} \& \bibinfo{author}{Suter, D.}
\newblock \bibinfo{title}{Measuring the spectrum of colored noise by dynamical
  decoupling}.
\newblock \emph{\bibinfo{journal}{Phys. Rev. Lett.}}
  \textbf{\bibinfo{volume}{107}}, \bibinfo{pages}{230501}
  (\bibinfo{year}{2011}).
\newblock
  \urlprefix\url{https://link.aps.org/doi/10.1103/PhysRevLett.107.230501}.

\bibitem{Gossard2012}
\bibinfo{author}{Medford, J.} \emph{et~al.}
\newblock \bibinfo{title}{Scaling of dynamical decoupling for spin qubits}.
\newblock \emph{\bibinfo{journal}{Phys. Rev. Lett.}}
  \textbf{\bibinfo{volume}{108}}, \bibinfo{pages}{086802}
  (\bibinfo{year}{2012}).
\newblock
  \urlprefix\url{https://link.aps.org/doi/10.1103/PhysRevLett.108.086802}.

\bibitem{Philips2022}
\bibinfo{author}{Philips, S. G.~J.} \emph{et~al.}
\newblock \bibinfo{title}{{Universal control of a six-qubit quantum processor
  in silicon}}.
\newblock \emph{\bibinfo{journal}{Nature}} \textbf{\bibinfo{volume}{609}},
  \bibinfo{pages}{919--924} (\bibinfo{year}{2022}).
\newblock \urlprefix\url{https://doi.org/10.1038/s41586-022-05117-x}.

\bibitem{Undseth2023}
\bibinfo{author}{Undseth, B.} \emph{et~al.}
\newblock \bibinfo{title}{Hotter is easier: Unexpected temperature dependence
  of spin qubit frequencies}.
\newblock \emph{\bibinfo{journal}{Phys. Rev. X}} \textbf{\bibinfo{volume}{13}},
  \bibinfo{pages}{041015} (\bibinfo{year}{2023}).
\newblock \urlprefix\url{https://link.aps.org/doi/10.1103/PhysRevX.13.041015}.

\end{thebibliography}

\clearpage

\section*{Extended Data}

\setcounter{figure}{0}
\setcounter{table}{0}

\begin{figure*}
  \centering
  \includegraphics[scale=0.9]{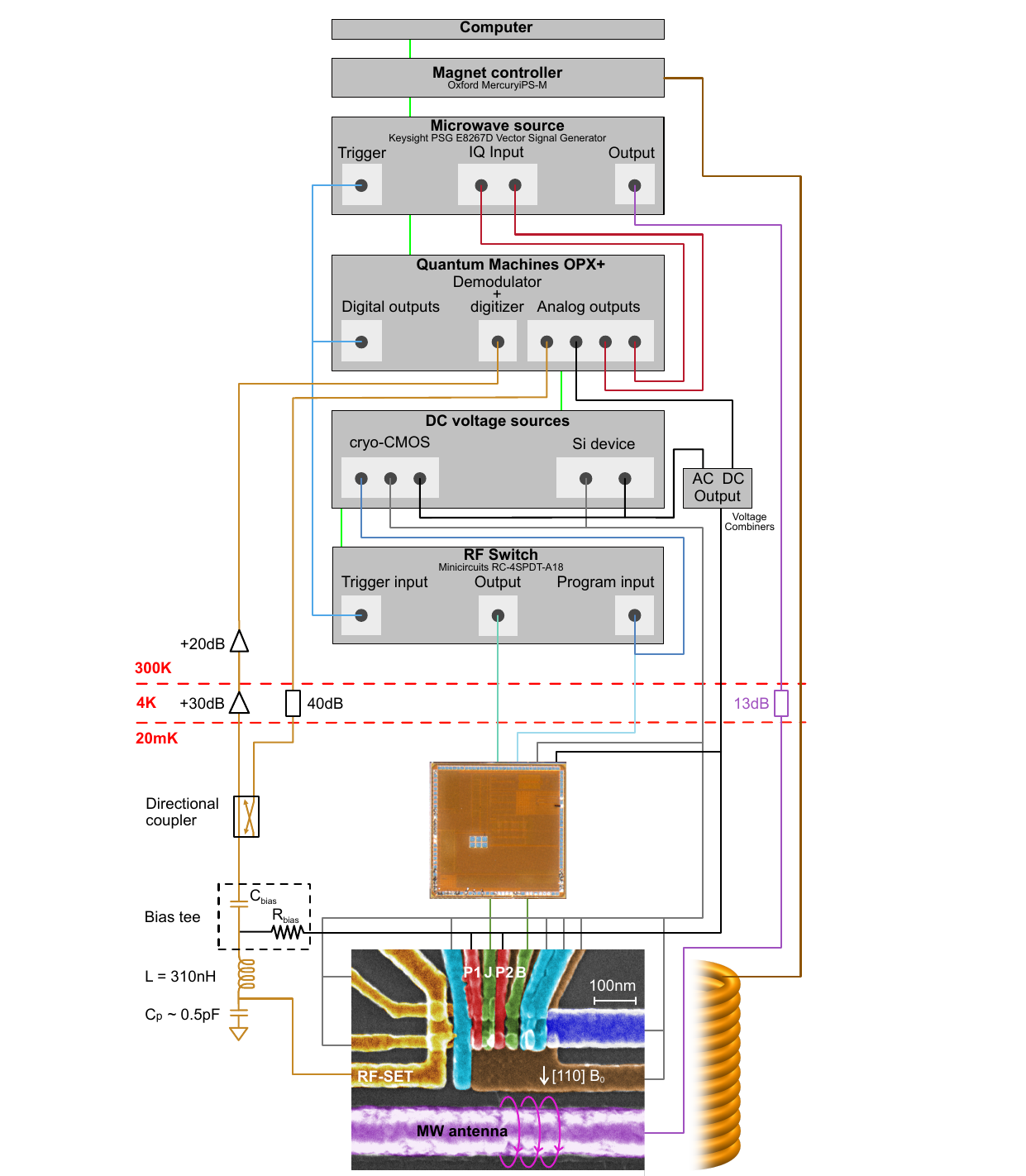}
  \caption{\textbf{Full experimental setup.}}
  \label{E1}
\end{figure*}

\begin{figure*}
\centering
  \includegraphics[scale=0.85]{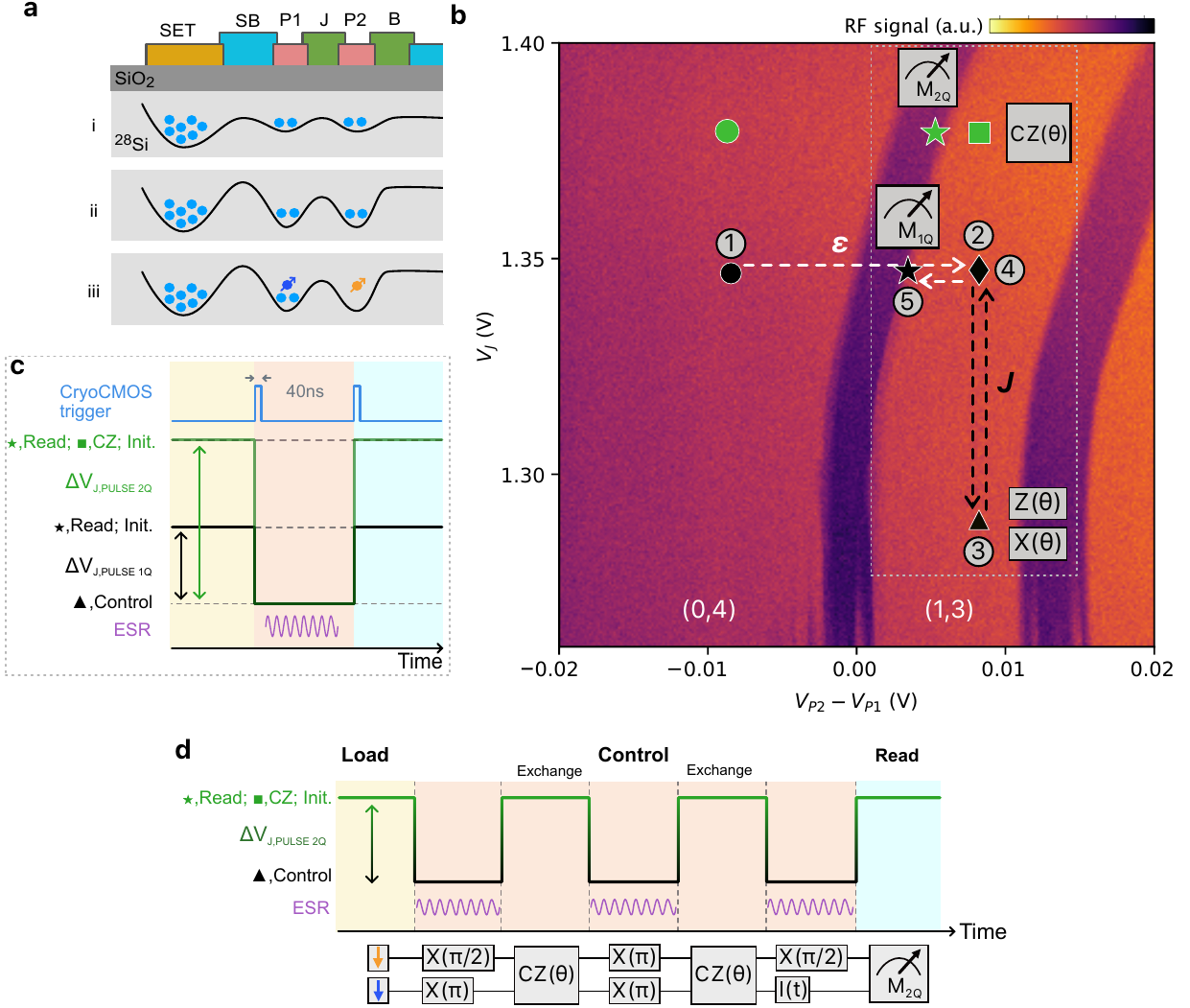}
  \caption{\textbf{Loading and pulsing scheme.} 
  \textbf{a}, the barrier to the SET reservoir is lowered for a set number of electrons to occupy the dots (i). This barrier is raised to isolate the dots (ii), and the electron states are initialized via tuning of the plunger gates P1 and P2 (iii). 
  \textbf{b}, Single-qubit pulsing scheme for ESR measurements. A diabatic detuning ramp $(1\rightarrow2)$ from the (0,4) to (1,3) state initialises the double quantum dot into a singlet state. The qubit is then pulsed using the $J$-gate to the control point $(2\rightarrow3)$ at which point a MW pulse is applied that rotates the target qubit in resonance with the MW frequency. The qubit is pulsed back to $\blacklozenge$ $(3\rightarrow4)$, then pulsed into the Pauli Spin Blockade regime $(4\rightarrow5)$ for readout. Detuning $\epsilon$ pulses use gates $P1$ and $P2$ and are always RT-operated. $J$-gate pulses are generated either at RT or by cryo-CMOS. The approximate operation points of single qubit (X, Z), two qubit (CZ), and readout points ($M_{1Q}$, $M_{2Q}$) are indicated by a triangle ($\blacktriangle$), square ($\blacksquare$), star ($\star$) and diamond ($\blacklozenge$) respectively. 
  \textbf{c} (inset of b), basic $J$-gate pulsing scheme for ESR measurements when using cryo-CMOS. Here, the gate $J$ is in a ``charge-locked" state held at $V_{HOLD}$. An external trigger is used to pulse between the two $J$-gate levels, whose separation is determined by $\Delta V_{J,PULSE}$. 
  \textbf{d}, DCZ sequence focussing on $J$-gate pulses. The qubits are initialised into a $T_-$ state, before $X$ and $X / 2$ gates are performed on the control and target qubits respectively. A spin-echo sequence is inserted between the exchange gates, which precedes the final projection for measurement of the two-qubit spin state.}
  \label{E2}
\end{figure*}

\begin{figure*}
\centering
  \includegraphics[scale=0.85]{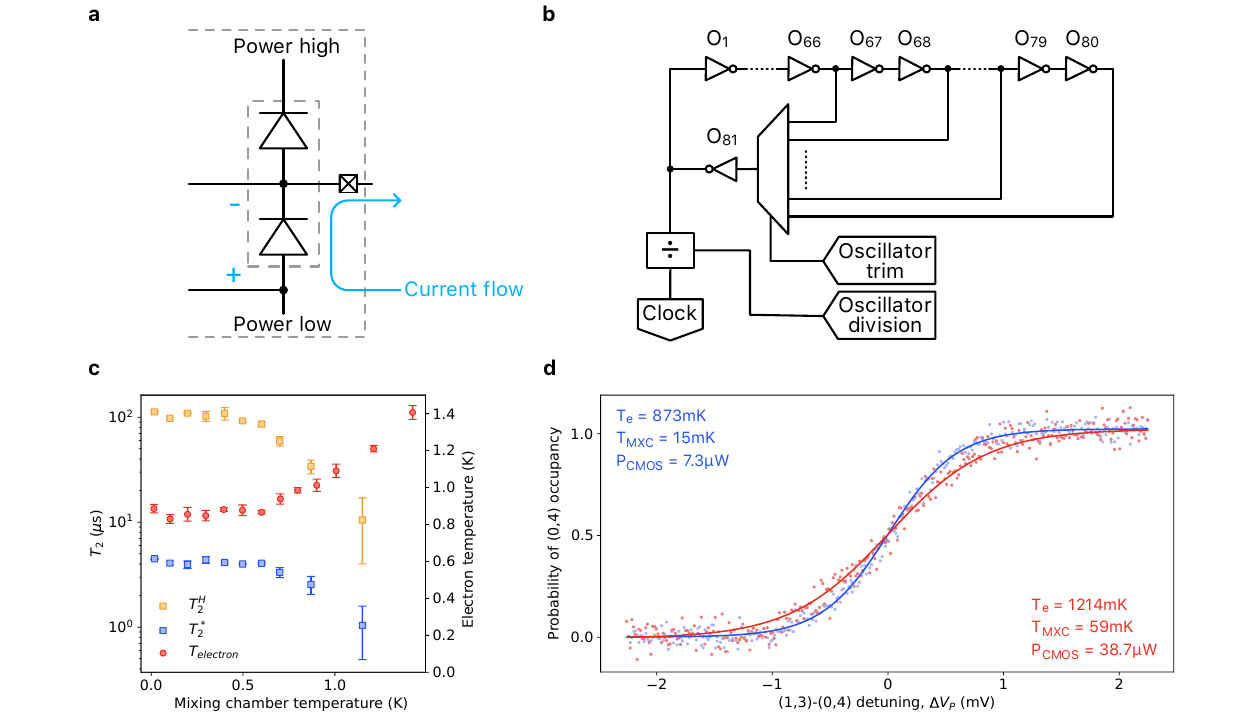}
  \caption{\textbf{Temperature dependence using artificial cryo-CMOS and mixing chamber heating.} 
  \textbf{a}, By forward biasing an ESD protection diode on-chip, the power draw can be programmed to mimic that of any oscillator trim or division value upon oscillator activation. Artificial power draw also allows for higher resolution investigation versus other parameters, as well as extrapolation beyond what the maximal cryo-CMOS power draw possible with it's feature set.
  \textbf{b}, Schematic of the cryo-CMOS ring oscillator. 81 in-series inverters are connected to a three-bit multiplexer, controlling the tap-off point. The frequency is tunable by programmable inputs into the oscillator trim. This frequency is further divided by the oscillator divider, which is eight-bit tunable. The ultimate output frequency is then passed on to the FSM.
  \textbf{c}, Hahn echo coherence time and Ramsey coherence time of Q1, and measured electron temperature as a function of mixing chamber temperature. Base effective electron temperature is approximately 850~mK, which only starts to increase once the mixing chamber exceeds 700~mK. Electron temperature and mixing chamber temperature equalize at 1~K
  \textbf{d} (1-3)-(0,4) charge occupation probability. Solid lines are Fermi distribution fits, allowing for extraction of effective electron temperature. Fitting to distributions with high mixing chamber temperature (at which point becoming equal to that of the sample) allows for determination of the lever arm.}
  \label{E3}
\end{figure*}

\begin{figure*}
\centering
  \includegraphics[scale=1]{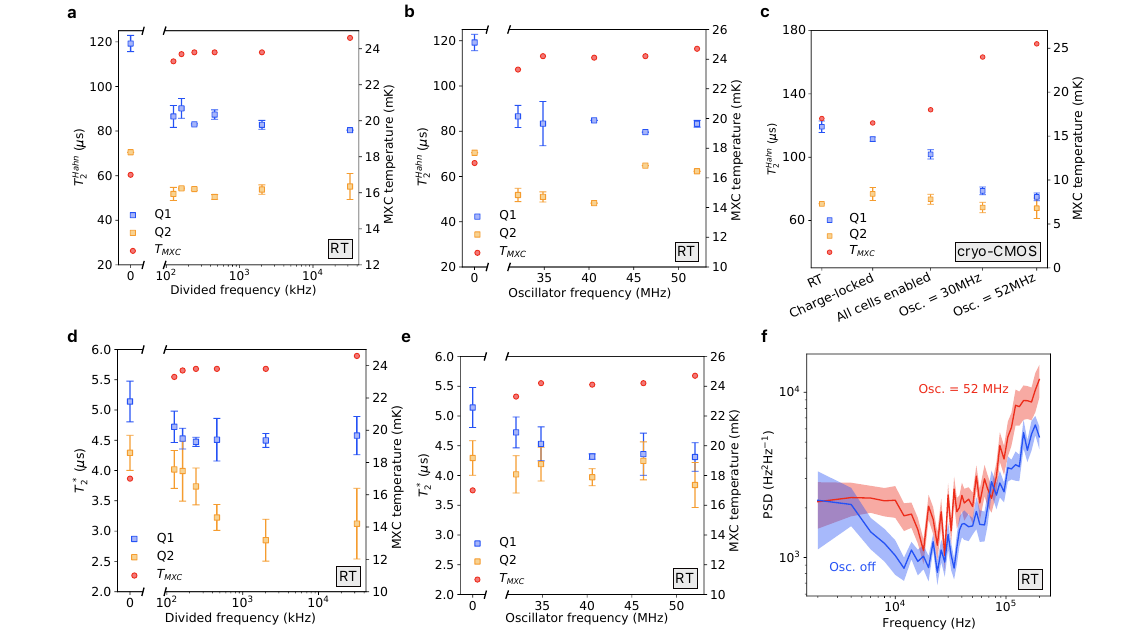}
  \caption{\textbf{Extended single qubit performance.} 
  \textbf{a}, $T_2^{Hahn}$ coherence of both qubits as a function of divided frequency from the cryo-CMOS oscillator. The output clock frequency (here set to 30~MHz) passed to the fast state machine is divided by an integer between 1 and 255, see Extended Data Fig.~\ref{E2} for schematic details. The inset indicates whether RT or cryo-CMOS control is used; all RT pulsing schemes are replicable by cryo-CMOS. Activating the oscillator immediately causes a drop in coherence due to the extra thermal dissipation, and lowering the divider value increases this dissipation slightly.
  \textbf{b}, Qubit coherence while directly changing oscillator frequency. The divider is set to a constant value of 255, and increasing oscillator frequency leads to an increase in power draw of the cryo-CMOS chip, reflected in the mixing chamber temperature.
  \textbf{c}, Similar to Fig.~\ref{Fig2}(c), $T_2^{Hahn}$ coherence is also explored as a function of various parameters under cryo-CMOS control conditions. 
  \textbf{d, e}, $T_2^*$ of both qubits while changing the divided and oscillator frequencies, similar to a and b.
  \textbf{f}, The noise power spectral density (PSD) is explored as a function of oscillator frequency. Noise spectroscopy, based on the Carr-Purcell-Meiboom-Gill (CPMG) protocol \cite{Sarma2007, Alvarez2011, Gossard2012}, uses a single qubit as a noise probe. A rise in the overall white noise level over the detectable frequency range when the oscillator is at its maximum frequency is observed. At higher frequencies, we observe an increase in PSD, likely due to high-power driving or miscalibration of microwave pulses \cite{Philips2022, Undseth2023, huang2023}. }
  \label{E4}
\end{figure*}

\begin{figure*}
\centering
  \includegraphics[scale=0.8]{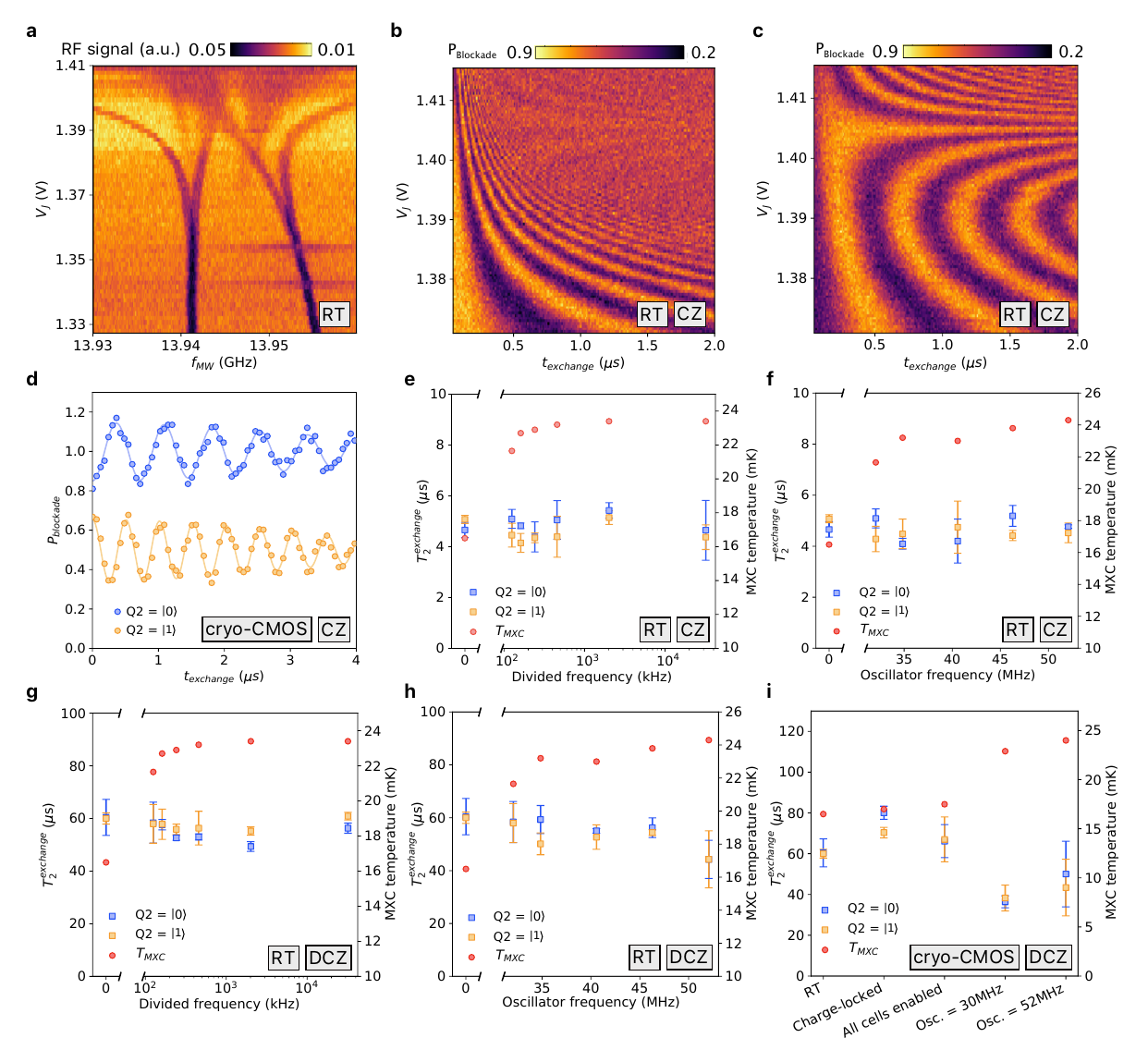}
  \caption{\textbf{Extended two-qubit performance and dependence on cryo-CMOS parameters.}
  \textbf{a}, ESR spectrum as a function of $V_J$. Exchange starts to open at around $V_J$ = 1.37~V. Measurements are done with room temperature $J$ pulses; RT or cryo-CMOS control is indicated in the lower right hand side of each figure.
  \textbf{b, c}, CZ oscillations as a function of exchange time $t_{exchange}$ and $V_J$. Multiple $J$ levels, not replicable by cryo-CMOS control are used here for optimal performance. The initial state of the target control qubits is indicated in the bottom right hand corner.
  \textbf{d}, CZ oscillations using cryo-CMOS control, showing the difference in visibility compared to \textbf{b-c}. Traces are offset for clarity. For all further figures, pulses are two-level and if generated at RT, are replicable by cryo-CMOS.
  \textbf{e}, similar to Extended Data Fig.~\ref{E4}(a), the two qubit CZ coherence time $T_2^{exchange}$ is explored as a function of divided frequency. The oscillator is set to 30~MHz for all divisions, and pulses are generated from room temperature.
  \textbf{f}, CZ coherence time $T_2^{exchange}$ is now explored as a function of oscillator frequency. The divider is set to 255 for all measurements. \textbf{g, h}, Measurements in \textbf{e, f} are replicated here, however now using a DCZ pulsing protocol, allowing for longer coherence times.
  \textbf{i}, DCZ coherence is explored using cryo-CMOS control under various parameters, similar to Fig.~\ref{Fig4}(d).}
  \label{E5}
\end{figure*}

\end{document}